Reply to the Comment by Poole et al. on "A tropical "NAT-like" belt observed from space"

V. Noel, LMD/IPSL, CNRS, France

H. Chepfer, LMD/IPSL, Univ. Paris 06, France

Corresponding author:

V. Noel

Laboratoire de Meteorologie Dynamique

Ecole Polytechnique

91128 Palaiseau

France

vincent.noel@lmd.polytechnique.fr

**Abstract**

In their comment, Poole et al. (2009) aim to show it is highly improbable that the observations described in Chepfer and Noel (2009), and described as "NAT-like" therein, are produced by Nitric Acid Trihydrate (NAT) particles. In this reply, we attempt to show why there is, in our opinion, too little evidence to reject this interpretation right away.

The main hypothesis in the Comment by Poole et al. (hereafter called P09) is that the optical observations labeled "NAT-like" in Chepfer and Noel 2009 (hereafter called CN09) are most likely a mixture of Liquid Aerosols (LA) and ice crystals, given particle number concentrations from the literature.

First of all, we want to emphasize that, contrarily to what might be inferred from reading P09, CN09 did not conclude that "NAT-like" observations should unambiguously be attributed to NAT particles; the term "NAT-like" referred to the optical behavior of the particles under study. Instead, CN09 proposed several possible interpretations, NAT particles being one of them. The present reply aims to show that, while P09 certainly brings cogent arguments that make the NAT interpretation less probable, it should not be dismissed entirely yet.

**1. On the detection of optically very thin layers with CALIOP**

P09 shows that NAT particle concentrations of $10^{-4}$ $cm^{-3}$ would produce a backscatter near $10^{-6}$ $km^{-1}$ $sr^{-1}$, too low to be observed from CALIOP.

CN09 used observations in the lower tropical stratosphere to calibrate the background backscatter (due to molecules), and identified layers as signal fluctuations above this background. By extracting the particulate backscatter this way, we found it was possible to detect layers with backscatter close to $10^{-5}$ $km^{-1}$ $sr^{-1}$ above the molecular signal in CALIOP data, i.e. to identify layers that backscatter less than the molecular background (3-5 $10^{-4}$ $km^{-1}$ $sr^{-1}$ as shown in P09). Those layers possess cloud-like qualities: horizontal and vertical extension, consistent geographic distribution over specific areas, homogeneous optical properties (at least as homogeneous as in PSC observations), etc.

A recent paper by Immler et al. (2007) shows a very similar layer observed by ground-based lidar, producing similar levels of backscatter (i.e. near $10^{-5}$ $km^{-1}$ $sr^{-1}$). The authors speculate this layer is a remnant of ice clouds, stabilized by inorganic acids, and conclude it is most likely made of NAT crystals in concentrations near 1 $L^{-1}$, i.e. $10^{-3}$ $cm^{-3}$. The concentration-backscatter relationship is consistent with results from P09, though both quantities are much higher. In our observations, NAT-like layers produce backscatters between $10^{-5}$ and 5 $10^{-4}$ $km^{-1}$ $sr^{-1}$.

Since the $10^{-4}$ $cm^{-3}$ concentration used in P09 for NAT is derived from in-situ studies (by definition limited in scope and altitude), the generality of these numbers cannot be assessed. Moreover, given the lower cut-size of the inlets used to bring particle-laden air into the instruments that measure $NO_y$, a significant amount of small NAT particles might be missed during in-situ measurements. Thus the existence of higher NAT concentrations (e.g. $10^{-3}$ $cm^{-3}$) cannot be discounted yet, for example close to lightning events where $HNO_3$ amounts increase. P09 conducted simulations of optical properties of particle mixtures with such enhanced NAT levels, but results from these computations only appear in their Fig. 2 and are somewhat surprising (see next section). Since their Fig. 1, which shows backscatter levels expected for various particle mixtures, does not include results from enhanced NAT levels, it does not preclude the possibility to detect NAT layers in such concentrations in CALIOP observations.

## 2. On the optical properties of particle mixtures

First, P09 argues that mixtures of LA and NAT particles produce depolarization ratio below the limit used in CN09 (0.05) to discriminate between liquid and solid particles: their Fig. 2 shows that, even considering enhanced levels of NAT ($10^{-3}$ $cm^{-3}$ as discussed in the previous section), the produced depolarization ratios do not get higher than 0.02. However, in Fig. 2 the enhanced NAT mixture does not produce higher backscatter than the mixture with a $10^{-4}$ $cm^{-3}$ NAT concentration ($0.3 < 1-1/R_{532} < 0.6$ in both cases), backscatter levels even appear to be lower with the enhanced mixture (near $1-1/R_{532} \sim 0.6$). It is rather surprising that increasing particle concentrations by an order of magnitude would not lead to increased backscatter. This result, and thus the depolarization ratio values, should be investigated further.

Second, P09 argues that a local mixture of liquid aerosols and ice crystals would lead to intermediate values of volumic backscatter and depolarization (between those produced by LA and ice crystals). From a theoretical point of view, it seems to us that the shape hypothesis done for ice and NAT particles is a key point. P09 used oblate particles for both, and it is unclear how much these shapes depolarize compared to others, especially at the considered particle sizes. A higher amount of oblate particles might then be needed to produce a given depolarization ratio compared to non-spheroidal particles - the dependence of depolarization ratio on particle shape is a well-known complex problem (e.g. Takano and Liou 1989). As in Fig. 2 of P09 the depolarization is used as a strong discrimination criteria, it may be useful to test others particle shapes to check how they affect the depolarization-to-concentration relationship. From an observational point of view, in CN09 we attempted to extract the particulate backscatter and depolarization from the volumic CALIOP observations by removing the background contribution. This approach was able to retrieve particulate depolarization ratios typical of ice crystals (0.4-0.6) in layers with a much lower volumic depolarization ratio (0-0.2), so we assumed the background depolarization (including LA) was correctly cleaned up. After this processing, the layers labeled NAT-like in CN09 still show a intermediate depolarization (0.1-0.3), while other layers are clearly gathered near 0 (liquid aerosols) or 0.4-0.6 (ice crystals).

Finally, regarding the optical properties of particle layers, in addition to the backscatter and depolarization ratios CN09 also used the color ratio, which has not been tested in P09. It maybe be useful to test whether the mixing of ice+liquid aerosols as proposed by P09 is consistent with observed color ratios.

Nonetheless, we agree that CN09 NAT-like layers could be due to (1) intense, small-scale, local increases of BSA concentrations mixed with ice crystals (in which case the process mentioned above would fail to totally remove the background contribution to the backscatter); however this hypothesis does not seem necessarily more supported by evidence than NAT particles, (2) more "exotic" processes (e.g. deposition of liquid aerosols on ice crystals leading to particle rounding and weakened depolarization). In those cases, the final particulate depolarization (after background correction) would indeed be intermediate and could be consistent with observations. It would be very interesting to find a way to discriminate between these two hypothesis (NAT particles / ice + liquid aerosols) based on optical observations alone.

## 3. Conclusion

As already stated in CN09, we agree with P09 that there is not enough evidence to unambiguously conclude that NAT-like layers in CN09 are actually composed of NAT crystals. Mixtures of liquid aerosols and ice, as proposed by P09, appear as a convincing possibility. As mentioned in CN09, other potential candidates include very small, non-spherical ice crystals that have been observed in situ in significant concentrations (Gayet et al. 2007); it remains to be seen if such particles are thermodynamically stable in the specific conditions of the TTL. However, this reply tries to show why we also think that it is too soon to definitely rule out a NAT composition for the non-ice / non-liquid layers in CALIOP TTL observations. More studies, including remote sensing and in-situ observations, are required to fully resolve these questions.